\documentclass[a4paper,11pt]{article}

\usepackage{amsmath}
\usepackage{amssymb}
\usepackage[dvips]{graphicx}
\usepackage[dvips]{psfrag}
\usepackage{cite}

\makeatletter
\@addtoreset{equation}{section}
\renewcommand{\theequation}{\thesection.\@arabic\c@equation}
\makeatother

\makeatletter
\renewcommand\appendix{\par
  \setcounter{section}{0}%
  \setcounter{subsection}{0}%
  \gdef\thesection{Appendix \@Alph\c@section }
  \renewcommand{\theequation}
  {\Alph{section}.\arabic{equation}}
}
\makeatother

\def \be {\begin{equation}}
\def \ee {\end{equation}}
\def \ba {\begin{array}}
\def \ea {\end{array}}
\def \bea{\begin{eqnarray}}
\def \eea{\end{eqnarray}}

\def \a {\alpha}
\def \b {\beta}
\def \g {\gamma}
\def \G {\Gamma}
\def \d {\delta}
\def \D {\Delta}
\def \e {\epsilon}

\def \m {\mu}
\def \n {\nu}

\def \l {\lambda}
\def \L {\Lambda}
\def \s {\sigma}
\def \S {\Sigma}
\def \r {\rho}
\def \o {\omega}
\def \O {\Omega}
\def \th {\theta}

\def \z {\zeta}

\def \p {\partial}
\def \f {\frac}
\def \na {\nabla}

\def \nn {\nonumber}
\def \ma {\mathcal}

\def \mb {\mathbb}

\def \lt {\left}
\def \rt {\right}

\def \ra {\rightarrow}
\def \sr {\sqrt}
\def \td {\tilde}
\def \hs {\hspace}

\begin{document}

\titlepage

\vspace*{-15mm}
\baselineskip 10pt

\baselineskip 24pt
\vglue 10mm

\begin{center}
{\Large\bf Novel CFT Duals for Extreme Black Holes}

\vspace{8mm}

\baselineskip 18pt

\renewcommand{\thefootnote}{\fnsymbol{footnote}}

Bin Chen$^{1,2}$\footnote[1]{\,bchen01@pku.edu.cn}
and
Jia-ju Zhang$^1$\footnote[2]{\,jjzhang@pku.edu.cn}

\renewcommand{\thefootnote}{\arabic{footnote}}

\vspace{5mm}

{\it
$^{1}$Department of Physics, and State Key Laboratory of Nuclear Physics and Technology,
Peking University, Beijing 100871, P.R. China\\
\vspace{2mm}
$^{2}$Center for High Energy Physics, Peking University, Beijing 100871, P.R. China
}

\vspace{10mm}

\end{center}

\begin{abstract}

In this paper, we study the CFT duals for extreme black holes in the stretched horizon formalism.  We consider the extremal RN,  Kerr-Newman-AdS-dS, as well as the higher dimensional Kerr-AdS-dS black holes. In all these cases, we reproduce the well-established CFT duals. Actually we show that for stationary extreme black holes, the stretched horizon formalism always gives rise to the same dual CFT pictures as the ones suggested by ASG of corresponding near horizon geometries. Furthermore, we propose new CFT duals for 4D Kerr-Newman-AdS-dS and higher dimensional Kerr-AdS-dS black holes. We find that every dual CFT is defined with respect to a rotation in certain angular direction, along which the translation defines a $U(1)$ Killing symmetry. In the presence of two sets of $U(1)$ symmetry, the novel CFT duals are generated by the modular group $SL(2,\mb Z)$, and for $n$ sets of $U(1)$ symmetry there are general CFT duals generated by T-duality group $SL(n,\mb Z)$.
\end{abstract}

\baselineskip 18pt

\newpage


\section{Introduction}

The Kerr/CFT correspondence was first proposed by Guica $et~al$~\cite{Guica:2008mu}. With suitable falloff conditions on the metric perturbation, the asymptotic symmetry group (ASG) of near horizon geometry of extreme Kerr black hole (NHEK)~\cite{Bardeen:1999px} was found to form  a Witt algebra (Virasoro algebra without central charge) classically, whose quantum version gives the Virasoro algebra with a central charge $c_L=12J$ where $J$ is the angular momentum of Kerr black hole. From Frolov-Thorne vacuum, it was found that the dual CFT could have a non-vanishing left temperature $T_L=1/2\pi$. Then the Bekenstein-Hawking entropy $S_{BH}=2\pi J$ could be reproduced by Cardy's formula
\be \label{e3}  S=\f{\pi^2}{3}c_L T_L. \ee
Therefore, it was argued that an extreme Kerr black hole could be holographically described by an 2D chiral conformal field theory (CFT) with the central charge $c_L=12J$.

Soon after this proposal, the same analysis was generalized to many other extreme black holes.\footnote{See the nice review~\cite{Bredberg:2011hp} for complete references.} In particular, the Kerr/CFT correspondence for four-dimensional extreme Kerr-Newman-AdS-dS black hole was studied by Hartman $et~al$ in~\cite{Hartman:2008pb}. It was found that the dual CFT description became singular in the extremal RN limit, and another CFT description, the so-called Q-picture, had to be introduced. To obtain the Q-picture,  the metric should be uplifted to five dimension, thus the $U(1)$ symmetry of the electromagnetic field behaves like the symmetry of a rotation. In the end, for a four-dimensional extreme Kerr-Newman black hole, there can be two independent pictures, namely the J-picture  and the Q-picture, corresponding to the original $U(1)$ rotation symmetry  and $U(1)$ gauge symmetry respectively. The similar phenomenon that different $U(1)$ symmetry may lead to different dual CFT description, all of which gives correct Bekenstein-Hawking entropy, has been found in other higher dimensional extreme Kerr black holes\cite{Lu:2008jk}. It has been suspected that the different CFT descriptions could be related to each other.

One of the key issues in Kerr/CFT correspondence is to determine the central charges of the dual CFTs. Besides the set of boundary conditions suggested in~\cite{Guica:2008mu}, another set of consistent boundary conditions was proposed in~\cite{Matsuo:2009sj}, which may describe the fluctuations of the extreme background and suggest that there exist a right-moving sector in the dual CFT. Such possibility was investigated from the point of view of AdS$_2$ quantum gravity in~\cite{Castro:2009jf}. It turned out that for extreme Kerr black hole, the dual CFT has both left and right sectors with the same central charge, while the excitations in the right sector are suppressed in the extreme limit. However, there is still short of consistent boundary conditions giving both central charges simultaneously. See~\cite{Matsuo:2009yet,Rasmussen:2009ix,Rasmussen:2010sa,Matsuo:2010newlimit,{Chen:2011wt}} for various efforts on this issue.

Very recently in a remarkable paper~\cite{Carlip:2011ax} Carlip re-derived the Kerr/CFT correspondence for the extreme Kerr black hole. He worked on the ordinary metric of an extreme Kerr black hole without taking the near horizon limit. Moreover, he did the analysis in the stretched horizon formalism to avoid the singular behavior at the horizon. With a suitable constraint of the falloff condition of the diffeomorphism on the shift vector $N^i$ at the stretched horizon, he found that the surface deformation bracket could form a Witt algebra classically, and then the quantum version of the Poisson bracket of symmetry generator form the Virasoro algebra with the same central charge $c_L=12J$.
Similarly he read the temperature of dual CFT  from Frolov-Thorne vacuum. Therefore Carlip reproduced successfully the holographic dual picture of extreme Kerr black hole originally proposed in~\cite{Guica:2008mu}.
It would be interesting to apply Carlip's prescription to other extreme black holes and see if it may shed light on the issue of  multiple dual CFT descriptions of the same extreme black hole.

In this paper we develop the stretched horizon formalism and investigate the CFT duals of various extreme black holes. We reproduce the well-established dual CFT descriptions for many kinds of extreme black holes,  including Kerr, NHEK, BTZ, RN, Kerr-Newman-AdS-dS, as well as the higher dimensional Kerr-AdS-dS black holes. We prove that for  stationary extreme black holes, in the stretched horizon formalism, it does not matter whether we work on the extreme black hole itself or its near horizon geometry, if we take its near horizon geometry as an extreme black hole. Moreover, we prove that the stretched horizon formalism is actually equivalent to the prescription used in~\cite{Guica:2008mu}, which is based on the asymptotic symmetry analysis of the near horizon geometry. For the RN black hole and the Q-picture of Kerr-Newman-AdS-dS black hole, we should uplift the metrics to higher dimensions following~\cite{Hartman:2008pb,Chen:2010bsa}. This uplift allows us to treat the $U(1)$ gauge field as an angular rotational symmetry in higher dimension. For the black holes with  multi-angular momenta, we obtain not only the discrete dual CFT pictures corresponding to independent $U(1)$ symmetry, but also novel dual conformal pictures generated by the T-duality group $SL(n,\mb Z)$.

We organize the remaining parts of the paper as follows.
In Section~2 we develop the stretched horizon formalism in a more systematic way in four dimensions  such that it can be applied to the other cases than extreme Kerr. We find that in the stretched horizon formalism we may work on the extremal black hole geometries directly or their near horizon geometries instead, both of which lead to the same holographic dual CFT picture.
In Section~3, we generalize the stretched horizon formalism to higher dimensions. In the case that the extreme black hole has multiple $U(1)$ isometries, we find that for each $U(1)$, there is a corresponding CFT dual.  Furthermore, we show that when the black hole geometry has $n$ $U(1)$ symmetries, we can generate novel CFT duals with $SL(n,\mb Z)$ T-duality group.
In Section~4,  we investigate four-dimensional Kerr-Newman-AdS-dS black hole. We re-derive the J-picture and Q-picture. Moreover we propose novel CFT duals, based on  appropriate combination of two $U(1)$ rotation via modular $SL(2,\mb Z)$ transformation.
In Section~5 we discuss the different pictures of higher dimensional Kerr-AdS-dS black holes. We mainly work on the five-dimensional extreme Kerr black holes, but also discuss briefly higher dimensional cases.
In Section 6, we discuss that the stretched horizon formalism is equivalent to the asymptotic symmetry group method in finding the holographic duals.
We end with the conclusions and discussions in Section 7.

\section{Stretched horizon formalism}

In this section we develop the stretched horizon formalism proposed by Carlip to study more general extreme black holes. We suggest~\cite{Carlip:2011ax} for more detailed discussion.  In the calculation we just use the ordinary extremal black hole metric and do not take the near horizon limit, however we can show that the near horizon geometry can give the same result.

\subsection{Four-dimensional black hole}

We take the four-dimensional stationary back hole metric in an ADM form
\be \label{e1}
ds^2=-N^2dt^2+h_{rr}dr^2+h_{\th\th}d\th^2+h_{\phi\phi}(d\phi+N^\phi dt)^2,
\ee
with $g_{\m\n}$ being independent of $t,\phi$.
When the black hole is extremal we have
\bea
&&N=f_1(r,\th)(r-r_+), \nn\\
&&h_{rr}=\f{f_2(r,\th)^2}{(r-r_+)^2}, \nn\\
&&N^\phi=-\O_H+(r-r_+)f_3(r,\th).
\eea
Here $r=r_+$ is the horizon of the black hole, $\O_H$ is the horizon angular velocity and $f_1,f_2,f_3$ are smooth functions at the horizon.

The only non-vanishing components of the canonical momentum are
\bea
&&\pi^{\th\phi}=-\f{\sqrt{h}}{2N}h^{\th\th}\p_\th N^\phi,\nn\\
&&\pi^{r\phi}=-\f{\sqrt{h}}{2N}h^{r r}\p_r N^\phi.
\eea
On the horizon $\ma H$ the metric is singular. Thus we just consider a surface near the horizon, which is called the ``stretched horizon'' $\ma H_s$.  On $\ma H_s$ we have $N^\phi$ expanded as
\be
N^\phi=-\O_H +\e,
\ee
where
\be
\e=(r-r_+)f_3|_{r=r_+}, \hs{3ex} f_3|_{r=r_+}=\p_r N^\phi|_{r=r_+}.
\ee
Generally $\e$ will depend on $\th$, but we just suppose that it is independent of $\th$, which is correct in all the cases we will consider. We will do all the calculation on the stretched horizon before we take the limit that $\ma H_s$ approaches the true event horizon to get the final result.

Under the diffeomorphism generated by a vector
\be \xi^\m =\lt( \f{\xi^\perp}{N},\hat{\xi}^i-\f{\xi^\perp}{N}N^i \rt), \ee
we have
\be
\d_\xi N^i=\bar \p_t \hat{\xi}^i+\hat{\xi}^j\p_j N^i -N^2 h^{ij}\p_j \lt(\f{\xi^\perp}{N}\rt),
\ee
where
\be \bar\p_t=\p_t-N^i\p_i=\p_t+\O_H\p_\phi -\e \p_\phi. \ee
We just focus on the components $\hat\xi^r$, $\hat\xi^\phi$ and set $\xi^\perp=\hat\xi^\th=0$. To the metric~(\ref{e1}), the constraint
\bea \label{shift}
&&\d_\xi N^r=\ma O\lt[ (r-r_+)^2 \rt], \nn\\
&&\d_\xi N^\phi=\ma O\lt[ (r-r_+)^2 \rt],
\eea
requires that
\bea
&&\hat \xi^r=(r-r_+)\Phi'(\phi-\O_H t)+  \ma O\lt[ (r-r_+)^2 \rt], \nn\\
&&\hat \xi^\phi=\Phi(\phi-\O_H t)  +\ma O\lt[ (r-r_+)^2 \rt],
\eea
where $\Phi(x)$ is an arbitrary function of $x$. If we take the leading order terms and make the Fourier decomposition $\Phi_m(x)=-e^{-imx}$, we get the Witt algebra
\be
i \lt[ \xi_m,\xi_n \rt]_{LD}=(m-n)\xi_{m+n},
\ee
where $LD$ means Lie bracket.

The symmetry generator of canonical general relativity is
\be \label{e6} H[\xi]=\int d^3 x \lt( \xi^\perp\ma H+ \hat\xi^i\ma H_i  \rt), \ee
where $\ma H, \ma H^i$ are the Hamiltonian and momentum constraints respectively. In order to have the variation of $H[\xi]$ well defined, a boundary term $B[\xi]$ must be added. Then we have the Poisson bracket
\be \lt\{ H[\xi],H[\eta] \rt\} = H[\{\xi,\eta\}_{SD}] +K[\xi,\eta] ,\ee
where the surface deformation brackets are
\bea
&&\{\xi,\eta\}_{SD}^\perp=\hat \xi^i D_i\eta^\perp  -\hat \eta^i D_i\xi^\perp,  \nn\\
&&\{\xi,\eta\}_{SD}^i=\hat \xi^k D_k \hat\eta^i  -\hat \eta^k D_k \hat\xi^i
                     +h^{ik}(\xi^\perp D_k \eta^\perp-\eta^\perp D_k \xi^\perp).
\eea
In our case the surface deformation bracket is equivalent to the Lie bracket.
The central charge term can be calculated as
\bea
&&K[\xi,\eta]=\cdots-\f{1}{8\pi G} \int_{\ma H_s} d^2 x \f{\sqrt{\s}}{\sqrt h}n^k \lt( \hat\eta_k
                 \pi^{mn}D_m\hat\xi_n  -\hat\xi_k \pi^{mn}D_m\hat\eta_n \rt)\nn \\
&&\phantom{K[\xi,\eta]} =\cdots+\f{f \ma A}{32\pi^2G}\int^{2\pi}_0d\phi
                                                       (\p^2_\phi\hat\xi^\phi\p_\phi\hat\eta^\phi                                                         -\p^2_\phi\hat\eta^\phi\p_\phi\hat\xi^\phi),
\eea
where the ellipsis denotes the terms that make no contributions to the central charge, $\ma A$ is the horizon area and
\be f=\f{{f_2}f_3}{f_1}\left|_{r=r_+}.\right. \label{f}\ee
which is assumed to be $\th$ independent as well.
Then we have
\be
i\lt\{ H[\xi_m],H[\xi_n] \rt\}_{PB}=(m-n)H[\xi_{m+n}]
                                + \f{f\ma A}{8\pi G}m^3\d_{m+n} +\ma Bm\d_{m+n},
\ee
where $\ma B$ comes from some possible contributions from $\ma B[\xi_{m+n}]$ and other terms of $K[\xi_m,\xi_n]$. After making the substitute
\be
H[\xi_m]\rightarrow L_m  +\f{f\ma A}{16\pi G}\d_m +\f{\ma B}{2}\d_m,
\ee
and changing the Poisson bracket to quantum commutator as $\{,\}_{PB}\ra[,]/i$, we finally get
\be
[L_m,L_n]=(m-n)L_{m+n}+\f{f\ma A}{8\pi G}m(m^2-1)\d_{m+n},
\ee
from which we can read out the central charge
\be \label{e4}  c_L=\f{3f \ma A}{2\pi G} .\ee

In order to use the Cardy formula to compute the entropy, we have to determine the temperatures of dual CFT. Here we just follow the work of~\cite{Carlip:2011ax,Guica:2008mu,Lu:2008jk} and adopt the Frolov-Thorne vacuum. Quantum fields near the horizon can be expanded by $e^{-i\o t+im\phi}$. Note that on the stretched horizon $\ma H_s$ the relevant angular coordinate is $\td\phi=\phi-(\O_H-\e)t$, and then we have
\be e^{-i\o t+im\phi}=e^{-i n_R t-in_L\td\phi}, \ee
where
\be n_R=\o-m(\O_H-\e),~~n_L=-m. \ee
After identifying the Boltzmann factor
\be e^{-(\o-m\O_H)/T_H}=e^{-n_L/T_L-n_R/T_R}, \ee
we can read out the temperature as
\be T_R=T_H,~~T_L=\f{T_H}{\e} . \ee
For an extremal black hole the Hawking temperature is zero, but we can evaluate at the stretched horizon $\ma H_s$ as
\be \label{e2} T_H =\f{1}{2\pi\sqrt{h_{rr}}}\p_r N=\f{f_1}{2\pi{f_2}}(r-r_+).\ee
As $\ma H_s\ra\ma H$, we have the temperature
\be \label{e5}  T_R=0,~~T_L=\f{1}{2\pi f}. \ee
Finally using the Cardy formula we obtain the microscopic entropy
\be S=\f{\pi^2}{3}c_L T_L=\f{\ma A}{4G},\ee
in perfect match with the macroscopic Bekenstein-Hawking entropy.

It is remarkable that for four-dimensional stationary extreme black hole of the metric (\ref{e1}), the central charge and the temperature of its dual CFT depend only on the
information of the metric components at the horizon. More precisely, they depend on a constant $f$, which is given by the relation (\ref{f}) and has nothing to do with the metric component $h_{\th\th}$.

\subsection{Simple examples}

We will present the discussions on Kerr-Newman black hole and higher dimensional Kerr black holes in the next few sections, where we will address the issue if there exist more general dual pictures of these black holes. Here we just present the other two cases, one is an extremal RN black hole, while the other is an extremal three-dimensional BTZ black hole.

For the former case, as the ordinary four-dimensional RN black hole has no angular velocity, so we cannot use the prescription we discussed above directly. However we may uplift RN black hole to 5D with metric~\cite{Chen:2010bsa,{Garousi:2009zx}}
\be ds^2=-N^2dt^2+\f{dr^2}{N^2}+r^2d \O ^2_2+\ell_5^2 \lt( d\chi-\f{\sr{G}Q}{\ell_5 r}dt \rt)^2, \label{RN}\ee
where $\ell_5$ is the scale of the extra dimension and  is sometimes chosen to be the Planck length $\ell_5=\sr G$. Here we leave $\ell_5$ to be a free parameter. We have the five-dimensional Newton constant $G_5=2\pi\ell G_4$, and the horizon area $\ma A_5=2\pi\ell_5 \ma A_4$, with $G_4=G$ and $\ma A_4=\ma A$.
For extremal black hole we have
\be N^2=\f{(r-r_+)^2}{r^2}. \ee
Notice that the metric~(\ref{RN}) looks quite similar to~(\ref{e1}). In the above analysis, the metric component $g_{\th\th}$ does not play an essential role in deriving the symmetry group. Similarly, the metric components $r^2d\O^2_2$ do not play a key role neither. Here the important point is the $U(1)$ symmetry along $\xi$ generate a Virasoro algebra and lead to a non-vanishing central charge.

Similar to the analysis before, for the metric~(\ref{RN}) we find that
\be
f_1=\f{1}{r},~~~f_2=r,~~~f_3=\f{\sr{G}Q}{\ell_5 rr_+},
\ee
and
\be f=\f{\sr{G}Q}{\ell_5}. \ee
Then we have
\be
c_L=\f{6\sr{G}Q^3}{\ell_5},\hs{3ex}
T_L=\f{\ell_5}{2\pi\sr{G}Q},
\ee
which is in agreement with the result found in~\cite{Chen:2010bsa}.

There is a subtle point in the above calculation. In using~(\ref{e4}), both the horizon area $\ma A$ and the Newton constant $G$ are changed for uplifted metric, but they receive the same modification that cancels with each other
\be
\frac{\ma A_5}{G_5}=\frac{\ma A}{G}.
 \ee
The ratio of the area of the horizon to the Newton constant is proportional to the entropy of the original black hole, which will not change no matter whether the black hole is uplifted or not.

Although our derivation in the last subsection was done in four-dimension, it can be generalized to lower or higher dimensional cases easily. For lower three-dimensional black hole, the situation is easier than the four-dimensional ones, we just repeat the previous process by just deleting the $\th$ component and the results do not change.
For the three-dimensional extremal BTZ black hole~\cite{BTZ}, it is of the metric
\be
ds^2=-\f{(r^2-r_+^2)^2}{\ell^2r^2}dt^2+\f{\ell^2r^2}{(r^2-r_+^2)^2}dr^2
       +r^2\lt( d\phi -\f{r_+^2}{\ell r^2}  dt\rt)^2 ,
\ee
then we have
\be
f_1=\f{r+r_+}{\ell r},~~~f_2=\f{\ell r}{r+r_+},~~~f_3=\f{r+r_+}{\ell r^2},
\ee
and
\be f=\f{\ell}{2r_+}. \ee
The horizon area of extremal BTZ back hole is
\be \ma A=2\pi r_+ . \ee
Thus we get the correct central charge and temperature
\be c_L=\f{3\ell}{2G} , \hs{3ex}
 T_L=\f{r_+}{\pi \ell} . \ee
The central charge was actually first obtained by Brown and Henneaux by considering the asymptotic behavior of the metric at infinity~\cite{Brown:1986nw}.

\subsection{Near horizon geometry}

Let us check the example of extreme Kerr black hole to support our result. For a Kerr black hole, its metric takes the form
\be
ds^2=-\f{\r^2\D}{\S}dt^2+\f{\r^2}{\D}dr^2+\r^2d\th^2
        +\f{\S\sin^2\th}{\r^2}\lt( d\phi -\f{2GMra}{\S}dt \rt)^2,
\ee
where
\be
\r^2=r^2+a^2\cos^2\th,
~~~\D=r^2-2GMr+a^2,
~~~\S=(r^2+a^2)^2-\D a^2\sin^2\th.
\ee
The angular momentum is $J=aM$, and the horizon locates at
\be r_\pm=GM\pm\sr{G^2M^2-a^2}. \ee
For an extreme Kerr black hole,  we have $GM=a$ and
\be
f_1={\f{\r}{\sr A}},
~~~f_2=\r,
~~~f_3|_{r=r_+}=\f{1}{2r_+^2}.
\ee
such that $f=1$.
The horizon area is
\be \ma A=8\pi r_+^2=8\pi GJ. \ee
Then we get the correct result
\be c_L=12J, \hs{3ex}
T_L=\f{1}{2\pi}. \ee

The next example is NHEK geometry itself. Since the NHEK metric can be regarded as an extremal black holes with the horizon $r_+=0$~\cite{Chen:2010ni,Chen:2010fr}, we can use our prescription to the NHEK metric. The metric of NHEK geometry is
\be ds^2=2GJ\O^2\lt(  -r^2dt^2+\f{dr^2}{r^2}+d\th^2+\L^2(d\phi+rdt)^2 \rt), \ee
where
\be \O^2=\f{1+\cos^2\th}{2},~~~\L=\f{2\sin\th}{1+\cos^2\th}. \ee
We can see that
\be f_1=f_2=\sr{2GJ\O^2},~~~f_3=1, \ee
and so $f=1$. The horizon area is
\be \ma A=2\pi\int^\pi_0d\th 2GJ\O^2\L=8\pi GJ. \ee
Then we have identical result as the Kerr case
\be c_L=12J,~~~T_L=\f{1}{2\pi}. \ee
This is exactly the result found in~\cite{Guica:2008mu}. Therefore we see that in the stretched horizon formalism, both the ordinary geometry of extremal Kerr black hole without taking near horizon limit and NHEK geometry taken as an extremal black hole give rise to same dual CFT description. Acutally this is true for all the extreme black holes of the 4D metric~(\ref{e1}).

For the 4D case, to get the near horizon  geometry of the stationary extreme black hole,  we  make the coordinate transformation on the metric~(\ref{e1})
\bea
t &\ra& \f{f_4}{\l r_+}t,  \nn\\
r &\ra& r_+(1+\l r),  \nn\\
\phi &\ra & \phi+\f{\O_H f_4}{\l r_+}t,
\eea
where $f_4=f_2/f_1|_{r=r_+}$  is a constant as we have assumed that $f$ and $f_3|_{r=r_+}$ are $\theta$ independent at the horizon. After taking the limit $\l\ra0$, we have the near horizon geometry whose metric is
\be \label{z1}
ds^2=\G(\th)\lt[ -r^2dt^2+\f{dr^2}{r^2}+\a(\th)d\th^2 \rt]+\g(\th)(d\phi+frdt)^2,
\ee
where
\be
\G(\th)=f_2^2|_{r=r_+},~~~
\G(\th)\a(\th)=h_{\th\th}|_{r=r_+},~~~
\g(\th)=h_{\phi\phi}|_{r=r_+},~~~
\ee
If we take the near horizon geometry as an extremal black hole with the horizon at $r=0$, then it can be justified easily that the metrics~(\ref{e1}) and~(\ref{z1}) give the same horizon area of black hole and the same central charge and temperature from the prescriptions~(\ref{e4}) and~(\ref{e5}).

\section{Higher dimensional black holes}

 For higher dimensional black holes, we take the five-dimensional black holes as the example to show that the above analysis still applies. We mainly focus the differences between the 4D and 5D cases and do not repeat too much the overlap part. Then we will show how to obtain novel CFT duals
 with modular group.

\subsection{General formalism}
We take the extreme 5D stationary back hole metric in an ADM form
\be \label{a1}
ds^2=-N^2dt^2+h_{rr}dr^2+h_{\th\th}d\th^2+\l_{mn}(d\phi^m+N^mdt)(d\phi^n+N^ndt),
\ee
with $\phi^m=(\phi,\chi)$, $N^m=(N^\phi,N^\chi)$ and
\be
\l_{mn}=\lt(\ba{cc}
h_{\phi\phi}& h_{\phi\chi} \\
h_{\chi\phi} & h_{\chi\chi}
\ea\rt).
\ee
We require that the black hole has two $U(1)$ isometries, corresponding to the rotations in $\phi$ and $\chi$. Therefore we have $g_{\m\n}$ being independent of $t,~\phi,~\chi$. At the stretched horizon, $N^\phi$ and $N^\chi$ are expanded as
\be
N^\phi=-\O_H^\phi+\e^\phi,\hs{3ex}N^\chi=-\O_H^\chi+\e^\chi,
\ee
where $(\O^\phi_H,\O^\chi_H)$ are two horizon angular velocities with respect to two rotating angles $(\phi,\chi)$ respectively, and
\bea
&&
~~~ \e^\phi=f_3^\phi|_{r=r_+}(r-r_+),
~~~ f_3^\phi|_{r=r_+}=\p_rN^\phi|_{r=r_+},\label{fphi}\\
&&
~~~ \e^\chi=f_3^\chi|_{r=r_+}(r-r_+),
~~~ f_3^\chi|_{r=r_+}=\p_rN^\chi|_{r=r_+}.\label{fchi}
\eea
The non-vanishing components of the canonical momentum now are
\bea
&& \pi^{\th\phi}=-\f{\sqrt{h}}{2N}h^{\th\th}\p_\th N^\phi,
~~~ \pi^{r\phi}=-\f{\sqrt{h}}{2N}h^{r r}\p_r N^\phi,   \nn\\
&& \pi^{\th\chi}=-\f{\sqrt{h}}{2N}h^{\th\th}\p_\th N^\chi,
~~~ \pi^{r\chi}=-\f{\sqrt{h}}{2N}h^{r r}\p_r N^\chi,   \nn
\eea

The diffeomorphism operating on the shift vectors becomes
\be
\d_\xi N^i=\bar \p_t \hat{\xi}^i+\hat{\xi}^j\p_j N^i -N^2 h^{ij}\p_j \lt(\f{\xi^\perp}{N}\rt),
\ee
with
\be \bar\p_t=\p_t-N^i\p_i=\p_t+\O_H^\phi\p_\phi+\O_H^\chi\p_\chi -\e^\phi \p_\phi -\e^\chi \p_\chi. \ee

As the black hole has two $U(1)$ rotational symmetries, we may find two dual CFT's from different points of view.
We define the first one with respect to the rotation in $\phi$ direction, and  call it the $\phi$-picture. In this case, we can focus on the
components $\hat\xi^r$, $\hat\xi^\phi$ and set $\xi^\perp=\hat\xi^\th=\hat\xi^\chi=0$. To the metric~(\ref{a1}), the constraint
\bea
&&\d_\xi N^r=\ma O\lt[ (r-r_+)^2 \rt], \nn\\
&&\d_\xi N^\phi=\ma O\lt[ (r-r_+)^2 \rt],
\eea
requires that
\bea
&&\hat \xi^r_{(\phi)}=(r-r_+)\Phi'(\phi-\O_H^\phi t)+  \ma O\lt[ (r-r_+)^2 \rt], \nn\\
&&\hat \xi^\phi_{(\phi)}=\Phi(\phi-\O_H^\phi t)  +\ma O\lt[ (r-r_+)^2 \rt],
\eea
where $\Phi(x)$ is an arbitrary function of $x$. We take the leading order terms and make the Fourier expansion $\Phi_m(x)=-e^{-imx}$, we get the Witt algebra
\be
i \lt[ \xi_m^{(\phi)},\xi_n^{(\phi)} \rt]_{LD}=(m-n)\xi_{m+n}^{(\phi)}.
\ee
Then we have the central charge term
\bea
&&K[\xi^{(\phi)},\eta^{(\phi)}]=\cdots-\f{1}{8\pi G} \int_{\ma H_s} d^3 x \f{\sqrt{\s}}{\sqrt h}n^k \lt( \hat\eta_k^{(\phi)}
                 \pi^{mn}D_m\hat\xi_n^{(\phi)}  -\hat\xi_k^{(\phi)} \pi^{mn}D_m\hat\eta_n^{(\phi)} \rt)\nn \\
&&\phantom{K[\xi^{(\phi)},\eta^{(\phi)}]} =\cdots+\f{f^\phi \ma A}{32\pi^2G}\int^{2\pi}_0d\phi
                               (\p^2_\phi\hat\xi^\phi_{(\phi)}\p_\phi\hat\eta^\phi_{(\phi)}                                                         -\p^2_\phi\hat\eta^\phi_{(\phi)}\p_\phi\hat\xi^\phi_{(\phi)}),
\eea
where
\be f^\phi=\f{{f_2}f_3^\phi}{f_1}\left|_{r=r_+}.\right. \ee
which is assumed to be $\th$ independent as well. We can read out the  central charge in the $\phi$-picture
\be \label{e19} c_L^\phi=\f{3f^\phi \ma A}{2\pi G} .\ee
Note that the central charge is proportional to the value of the function $f^\phi_3|_{r=r_+}$ in (\ref{fphi}), which is only related to the expansion of $N^\phi$
at the horizon and has nothing to do with $N^\chi$ or $\l_{mn}$.

On the other hand, we may define the second dual CFT with respect to the rotation in $\chi$ direction, and call it the $\chi$-picture.
Then we just focus on the components $\hat\xi^r$, $\hat\xi^\chi$ and set $\xi^\perp=\hat\xi^\th=\hat\xi^\phi=0$. The constraint
\bea
&&\d_\xi N^r=\ma O\lt[ (r-r_+)^2 \rt], \nn\\
&&\d_\xi N^\chi=\ma O\lt[ (r-r_+)^2 \rt],
\eea
requires that
\bea
&&\hat \xi^r_{(\chi)}=(r-r_+)X'(\chi-\O_H^\chi t)+  \ma O\lt[ (r-r_+)^2 \rt], \nn\\
&&\hat \xi^\chi_{(\chi)}=X(\chi-\O_H^\chi t)  +\ma O\lt[ (r-r_+)^2 \rt],
\eea
where $X(x)$ is an arbitrary function of $x$. We take the leading order terms and make the Fourier expansion $X_m(x)=-e^{-imx}$, we get the Witt algebra
\be
i \lt[ \xi_m^{(\chi)},\xi_n^{(\chi)} \rt]_{LD}=(m-n)\xi_{m+n}^{(\chi)}.
\ee
Note that the two Witt algebras are commutable,
\be
\lt[ \xi_m^{(\phi)},\xi_n^{(\chi)} \rt]_{LD}=0.
\ee
Then the central charge term becomes
\bea
&&K[\xi^{(\chi)},\eta^{(\chi)}]=\cdots-\f{1}{8\pi G} \int_{\ma H_s} d^3 x \f{\sqrt{\s}}{\sqrt h}n^k \lt( \hat\eta_k^{(\chi)}
                 \pi^{mn}D_m\hat\xi_n^{(\chi)}  -\hat\xi_k^{(\chi)} \pi^{mn}D_m\hat\eta_n^{(\chi)} \rt)\nn \\
&&\phantom{K[\xi^{(\chi)},\eta^{(\chi)}]} =\cdots+\f{f^\chi \ma A}{32\pi^2G}\int^{2\pi}_0d\chi
                               (\p^2_\chi\hat\xi^\chi_{(\chi)}\p_\chi\hat\eta^\chi_{(\chi)}                                                         -\p^2_\chi\hat\eta^\chi_{(\chi)}\p_\chi\hat\xi^\chi_{(\chi)}),
\eea
where
\be f^\chi=\f{{f_2}f_3^\chi}{f_1}\left|_{r=r_+}.\right. \ee
which is assumed to be $\th$ independent as well. We can read out the  central charge in the $\chi$-picture
\be \label{e20} c_L^\chi=\f{3f^\chi \ma A}{2\pi G} .\ee
Similarly, the central charge is proportional to the value of the function $f^\chi_3|_{r=r_+}$ in (\ref{fchi}), which depends only on $N^\chi$.

For the 5D case we can also use the Frolov-Thorne vacuum to determine the CFT temperature. Quantum fields near the horizon can be expanded by $e^{-i\o t+im\phi+ie\chi}$. In the $\phi$-picture, we just focus on the $\phi$ direction by turning off the $\chi$ direction mode $e=0$. At the stretched horizon $\ma H_s$ the relevant angular coordinate is $\td\phi=\phi-(\O_H^\phi-\e^\phi)t$, and then we have
\be e^{-i\o t+im\phi}=e^{-i n_R t-in_L^\phi\td\phi}, \ee
where
\be n_R=\o-m(\O_H^\phi-\e^\phi),~~~n_L^\phi=-m. \ee
After identifying the Boltzmann factor
\be e^{-(\o-m\O_H^\phi)/T_H}=e^{-n_R/T_R-n_L^\phi/T_L^\phi}, \ee
we can read out the temperature as
\be
T_R=T_H,~~~T_L^\phi=\f{T_H}{\e^\phi}.
\ee
Using the Hawking temperature~(\ref{e2}), as $\ma H_s\ra\ma H$, we have the $\phi$-picture temperatures
\be \label{e21}  T_R=0,~~~T_L^\phi=\f{1}{2\pi f^\phi}. \ee
Similarly, in the $\chi$-picture we can set the $\phi$ direction mode $m=0$, and get the temperature
\be \label{e32} T_L^\chi=\f{1}{2\pi f^\chi}. \ee
Finally using the Cardy formula, we obtain the microscopic entropy
\be S=\f{\pi^2}{3}c_L^\phi T_L^\phi=\f{\pi^2}{3}c_L^\chi T_L^\chi=\f{\ma A}{4G},\ee
in perfect match with the macroscopic Bekenstein-Hawking entropy. Note that we now have two different CFT duals, which we call the $\phi$-picture and the $\chi$-picture respectively, with different central charges and temperatures.

The above discussion could be applied to the other higher dimensional extreme black holes with multi-$U(1)$ isometries. These black holes include higher dimensional Kerr black holes in AdS and dS, and the black holes with multi-$U(1)$ charges.  Once the metrics of these black holes could be put into a form like (\ref{a1}), the discussion is straightforward. It could be expected that in these cases, there are multiple holographic CFT duals. In the following, we will show that there are actually more novel CFT duals generated by the T-duality group.

For the 5D case, we make the following coordinate transformation on the metric~(\ref{a1})
\bea
t &\ra& \f{f_4}{\l r_+}t,  \nn\\
r &\ra& r_+(1+\l r),  \nn\\
\phi &\ra & \phi+\f{\O_H^\phi f_4}{\l r_+}t, \nn\\
\chi &\ra & \chi+\f{\O_H^\chi f_4}{\l r_+}t
\eea
with $f_4=f_2/f_1|_{r=r_+}$ being a constant. Taking the limit $\l\ra0$, we have the near horizon geometry
\be \label{e22}
ds^2=\G(\th)\lt[ -r^2dt^2+\f{dr^2}{r^2}+\a(\th)d\th^2 \rt]+\g_{mn}(\th)(d\phi^m+f^mrdt)(d\phi^n+f^nrdt),
\ee
with
\bea
&&\G(\th)=f_2^2|_{r=r_+},~~~
\G(\th)\a(\th)=h_{\th\th}|_{r=r_+}, \nn\\
&&\g_{mn}(\th)=\l_{mn}|_{r=r_+},~~~
f^m=(f^\phi,f^\chi).
\eea
Like the 4D case, we take the near horizon geometry as an extremal black hole with the horizon at $r=0$, then the metrics~(\ref{a1}) and~(\ref{e22}) give the same horizon area of black hole and the same central charges and temperatures from the prescriptions~(\ref{e19}),~(\ref{e20}),~(\ref{e21}) and~(\ref{e32}).

\subsection{Novel CFT duals via T-duality}

As shown above, there are two independent CFT descriptions, depending on different $U(1)$ rotational symmetries. This inspires us to investigate other CFT duals, which depends on a generic $U(1)$ rotation. Naively, we may expect that the linear combination of two Killing symmetry  works. For example,
we can define  new angular coordinates $(\phi^\a,\phi^\b)$ as
\be
\lt(\ba{l} \phi^\a \\ \phi^\b \ea\rt)=
\lt(\ba{ll} \a & 1-\a \\ \b & 1-\b \ea\rt)
\lt(\ba{l} \phi \\ \chi \ea\rt), \label{ang1}
\ee
with $\a,\b\in[0,1]$ and $\a\neq\b$. Or as in~\cite{Chen:2009xja}  the new angular coordinates could be defined as
\be
\lt(\ba{c} \phi' \\  \chi' \ea \rt)=
\lt( \ba{cc} \cos\a & -\sin\a \\ \sin\a & \cos\a \ea \rt)
\lt(\ba{c} \phi \\  \chi \ea \rt),\label{ang2}
\ee
with $\a$ as the parameter. However, it turns out such coordinate transformations do not lead to well-defined CFT duals.
One essential requirement in the above treatment of 5D extreme black holes is that both angular variables should be periodic independently.
Obviously, the angular coordinates defined in (\ref{ang2}) are not periodic. Though the angular variables defined in (\ref{ang1}) seem to be periodic,
they are not periodic independently in the sense that for a fixed $\phi^\a$, $\phi^\b$ is not periodic, and vice versa.

To keep the periodicity, we need to use the T-duality group to define the new angular coordinates. For the 5D black holes with two $U(1)$
isometries, the group is just the modular group $SL(2,\mb Z)$ of the torus $T^2$. In fact, the last term of metric (\ref{e18}) can be realized as a torus and $SL(2,\mb Z)$ is the corresponding modular group. We propose the following coordinate transformation
\be
\lt(\ba{c} \phi' \\  \chi' \ea \rt)=
\lt( \ba{cc} a & b \\ c & d \ea \rt)
\lt(\ba{c} \phi \\  \chi \ea \rt),
\ee
with the parameters
\be
\lt( \ba{cc} a & b \\ c & d \ea \rt) \in SL(2, \mb Z).
\ee
Actually, as
 \be
\lt(\ba{l} \phi \\ \chi \ea\rt) \cong \lt(\ba{l} \phi \\ \chi \ea\rt)+2\pi\lt(\ba{l} m \\ n \ea\rt),
\ee
with $m,n$ being arbitrary integers, the new coordinates have the periodicity
\be
\lt(\ba{l} \phi' \\ \chi' \ea\rt) \cong \lt(\ba{l} \phi' \\ \chi' \ea\rt)+2\pi\lt(\ba{l} m' \\ n' \ea\rt),
\ee
with
\be
\lt(\ba{c} m' \\ n' \ea \rt)=
\lt( \ba{cc} a & b \\ c & d \ea \rt)
\lt(\ba{c} m \\  n \ea \rt),
\ee
being arbitrary integers as well. Thus $(\phi',\chi')$ have the same periodicity as $(\phi,\chi)$, and thus can be treated as new
well-defined angular coordinates. Such coordinate transformation has also been discussed in \cite{Loran:2009cr}, where different
dual pictures have been suggested.

We write $\phi^{m'}=(\phi',\chi')$, $N^{m'}=(N^{\phi'},N^{\chi'})$, the transformation matrix and its reverse as
\be
\L^{m'}_{\phantom{m'}m}=\lt( \ba{cc} a & b \\ c & d \ea \rt),
\hs{3ex}
\L^{m}_{\phantom{m}m'}=\lt( \ba{cc} d & -b \\ -c & a \ea \rt).
\ee
Then we can rewrite the metric (\ref{a1}) as
\be \label{e24}
ds^2=-N^2dt^2+h_{rr}dr^2+h_{\th\th}d\th^2+
        +\l_{m'n'}(d\phi^{m'}+N^{m'}dt)(d\phi^{n'}+N^{n'}dt),
\ee
with
\be
\phi^{m'}=\L^{m'}_{\phantom{m'}m}\phi^m, \hs{3ex}N^{m'}=\L^{m'}_{\phantom{m'}m}N^m,\hs{3ex}
\l_{m'n'}=\L^{m}_{\phantom{m}m'}\L^{n}_{\phantom{m}n'}\l_{mn}.
 \ee
 Since $(\phi',\chi')$ can be regarded as angular coordinates and the translations along them generate two $U(1)$ isometries, we may
 define new CFT duals with respect to them. As $\det\l_{m'n'}=\det\l_{mn}$, we see that the black hole horizon area, and hence the entropy, do not change under the $SL(2,\mb Z)$ coordinate transformation. The only modification comes from the function $N^{m'}$, which leads to the change of constant $f^{m'}$. Therefore we can read the novel CFT duals with the central charges and temperatures directly
\bea
c_L^{m'}=\L^{m'}_{\phantom{m'}m}c_L^m,\hs{3ex}\f{1}{T_L^{m'}}=\L^{m'}_{\phantom{m'}m}\f{1}{T_L^{m}}, \label{novel}
\eea
with $c_L^{m'}=(c_L^{\phi'},c_L^{\chi'})$, $c_L^{m}=(c_L^{\phi},c_L^{\chi})$, $T_L^{m'}=(T_L^{\phi'},T_L^{\chi'})$ and $T_L^{m}=(T_L^{\phi},T_L^{\chi})$. More explicitly, we have
\bea
&& c_L^{\phi'}=a c_L^\phi + b c_L^\chi, ~~~
T_L^{\phi'}=\f{1}{a/T_L^\phi+b/T_L^\chi},\nn\\
&&c_L^{\chi'}=c c_L^\phi + d c_L^\chi, ~~~
T_L^{\chi'}=\f{1}{c/T_L^\phi+d/T_L^\chi}.
\eea
Here we have two novel CFT dual pictures, the $\phi'$-picture and the $\chi'$-picture,  defined with respect to the angular direction $\phi'$ and $\chi'$. The general pictures are generated by the modular group $SL(2,\mb Z)$ acting on the original $\phi$- and $\chi$-pictures. It is easy to see that using Cardy's formula, we can exactly reproduce the Bekenstein-Hawking entropy.

We note our results disagree with that of \cite{Loran:2009cr}, though the same coordinate transformation has been applied. In \cite{Loran:2009cr},
to keep $SL(2,\mb Z)$ invariance, a set of looser boundary conditions and moreover a traceless condition on the fluctuations in the original coordinates were imposed to get the asymptotic
symmetry group, which contains a subgroup generated by Virasoro algebra. The central charge they obtained in the simplest case corresponds to the case $a=b=1$. However, for the
more general case, our result differs from theirs. From our point of view, the central charges and temperatures for general CFT dual in \cite{Loran:2009cr} is problematic. We believe a careful ASG analysis based on the metric (\ref{e24}) will give the same result as ours.
The equivalence of our treatment to ASG formalism will be given in Section 6.

The above discussion could be generalized to other black holes with more $U(1)$ symmetries. In general, if the metric of the black hole
could be put into the form like (\ref{a1}), with $n$ $U(1)$ isometries, then the coordinate transformation keeping the periodicity is just
\be
\phi^{m'}=\L^{m'}_{~~m}\phi^m
\ee
with $\L^{m'}_{~~m}$ being a $SL(n,\mb Z)$ matrix. With respect to new angular variables $\phi^{m'}$, we may define novel CFT duals
with the central charges and temperatures as (\ref{novel}).
Therefore, a general CFT dual could be generated by the group $SL(n,\mb Z)$, which is a subgroup of T-duality group $O(n,n,\mb Z)$.

\section{Kerr-Newman-AdS-dS black hole}

In this section we turn to the study of extremal Kerr-Newman-AdS-dS black holes. We firstly re-derive the J-picture and the Q-picture, and then propose new dual pictures corresponding to other $U(1)$ rotations. In the process we have to uplift the 4D metric to a 5D one, which is quite similar to the 5D Kerr black hole. We show that we may produce the same pictures from near horizon geometry as well. This suggests that even in higher dimensions, it make no difference whether we work on the black hole background or its near horizon geometry, as proved in Section 3.

\subsection{J-picture}

The metric of Kerr-Newman-AdS-dS black hole could be written in an ADM form~\cite{Caldarelli:1999xj}
\be \label{j2}
ds^2=-\f{\r^2\D_\th\D_r}{\S^2}dt^2+\f{\r^2}{\D_r}dr^2+\f{\r^2}{\D_\th}d\th^2
        +h_{\phi\phi}\lt( d\phi +N^\phi dt \rt)^2,
\ee
where
\bea
&&\r^2=r^2+a^2\cos^2\th,~~~\D_\th=1-\f{a^2}{\ell^2}\cos^2\th, ~~~\D_r=(r^2+a^2)(1+r^2/\ell^2)-2Mr+q^2, \nn\\
&&\S^2=(r^2+a^2)^2\D_\th-\D_r a^2\sin^2\th, ~~~h_{\phi\phi}=\f{\S^2\sin^2\th}{\r^2\Xi^2},  \nn\\
&&N^\phi=-\f{a\Xi}{\S^2}[\D_\th(r^2+a^2)-\D_r],~~~\Xi=1-a^2/\ell^2,~~~q^2=q_e^2+q_m^2. \nn
\eea
The constant $1/\ell^2$ is positive (negative) for AdS (dS), and is zero for asymptotic flat spacetime. For an extremal black hole at the horizon we have $\D_r|_{r=r_+}=\p_r\D_r|_{r=r_+}=0$, which gives
\be
M=\f{r_+\lt[ (1+r_+^2/\ell^2)^2-q^2/\ell^2 \rt]}{1-r_+^2/\ell^2},~~~
a^2=\f{r_+^2+3r_+^4/\ell^2-q^2}{1-r_+^2/\ell^2}.
\ee
We can also define the constant $V$ as
\be V=\f{1}{2}\p^2_r\D_r|_{r=r_+}=\f{1+6r_+^2/\ell^2-3r_+^4/\ell^4-q^2/\ell^2}{1-r_+^2/\ell^2}. \ee
The physical mass, angular momentum, and the electric and magnetic charges are respectively
\be
M_{\textrm{ADM}}=\f{M}{\Xi^2},~~~J=\f{aM}{\Xi^2},~~~Q_e=\f{q_e}{\Xi},~~~Q_m=\f{q_m}{\Xi}.
\ee
The gauge field is
\be
 A=-\f{q_er}{\r^2} \lt( dt-\f{a\sin^2\th}{\Xi}d\phi \rt) -\f{q_m\cos\th}{\r^2}(adt-\f{r^2+a^2}{\Xi}d\phi).
\ee
The horizon area is calculated to be
\bea
&&\ma A=2\pi\int^\pi_0d\th\sr{\f{\r^2 h_{\phi\phi}}{\D_\th}}=\f{4\pi(r_+^2+a^2)}{\Xi} \nn\\
&&\phantom{\ma A}=\f{4\pi(2r_+^2+2r_+^4/\ell^2-q^2)}{1-2r_+^2/\ell^2-3r_+^4/\ell^2+q^2/\ell^2}.
\eea

From the metric we read
\bea \label{e23}
&&f_1^J|_{r=r_+}=\f{\sr{\r_+^2V}}{r_+^2+a^2}, ~~~
f_2^J|_{r=r_+}=\sr{\f{\r_+^2}{V}}, ~~~
f_3^J|_{r=r_+}=\f{2a\Xi r_+}{(r_+^2+a^2)^2}, \nn\\
&&f^J=k=\f{2ar_+\Xi r_0^2}{(r_+^2+a^2)^2},
\eea
where
\bea
&& \r_+^2=r_+^2+a^2\cos^2\th,
~~~r_0^2=\f{(r_+^2+a^2)(1-r_+^2/\ell^2)}{1+6r_+^2/\ell^2-3r_+^4/\ell^4-q^2/\ell^2}.
\eea
Note that $f_1|_{r=r_+},f_2|_{r=r_+}$ are not $\th$ independent, but their ratio $f_2/f_1|_{r=r_+}$ is.
From~(\ref{e4}) and~(\ref{e5}), we obtain the central charge and the temperature directly
\bea \label{jp}
&&c_L^J=\f{6k(r_+^2+a^2)}{\Xi}=\f{12r_+\sr{(1-r_+^2/\ell^2)(r_+^2+3r_+^4/\ell^2-q^2)}}
            {1+6r_+^2/\ell^2-3r_+^4/\ell^4-q^2/\ell^2}, \\
&&T_L^J=\f{1}{2\pi k}  \nn\\
&&\phantom{T_L}=\f{(1+6r_+^2/\ell^2-3r_+^4/\ell^4-q^2/\ell^2)(2r_+^2+2r_+^4/\ell^2-q^2)}
       {4\pi r_+(1-2r_+^2/\ell^2-3r_+^4/\ell^2+q^2/\ell^2)
          \sr{(1-r_+^2/\ell^2)(r_+^2+3r_+^4/\ell^2-q^2)}},\nn
\eea
which are in perfect match with the result in~\cite{Hartman:2008pb,HiddenSymmetry,Chen:2010xu,Chen:2010bh}. In the limit $1/\ell\ra0$, this is just the well-known result
\be \label{jp1}
c_L^J=12J,~~~T_L^J=\f{r_+^2+a^2}{4\pi J}.
\ee
From above analysis, it is obvious that the J-picture is defined with respect to the rotation in $\phi$ direction.

In fact we are a little surprised to get the result for Kerr-Newman black hole, since the generator~(\ref{e6}) we used are got from the pure gravity, without considering the possible contributions from the gauge field. However it is argued in~\cite{Hartman:2008pb} that gauge field makes no contribution to the central charge, and $c=c_{\textrm{grav}}$. So our treatment focus only on the geometry, it seems suggest that the gauge field will not contribute to the central charge in any charged black hole cases.

\subsection{Q-picture}

In the limit $a\ra0$, $k$ tends to be zero and the J-picture shown in the last subsection becomes singular. One has to consider the so-called Q-picture, which may exist with arbitrary $a$ as well. In this case, the four-dimensional charged black hole metric can be uplifted to a five-dimensional uncharged black hole metric~\cite{Hartman:2008pb,Chen:2010bsa}
\be \label{j1}
ds_5^2=-\f{\r^2\D_\th\D_r}{\S^2}dt^2+\f{\r^2}{\D_r}dr^2+\f{\r^2}{\D_\th}d\th^2
        +h_{\phi\phi}\lt( d\phi +N^\phi dt \rt)^2 +(d\chi+A)^2,
\ee
where $\chi\sim\chi+2\pi$. Note that the gauge field $A$ can be decomposed as
\be
A=A_1 (d\phi+N^\phi dt)+N^\chi dt,
\ee
with
\bea
&& A_1=\f{q_e ra\sin^2\th+q_m(r^2+a^2)\cos\th}{\r^2\Xi}, \nn\\
&& N^\chi=-\f{q_e\D_\th r(r^2+a^2)+q_ma\D_r\cos\th}{\S^2}.
\eea
Then the metric~(\ref{j1}) can be recast as
\be \label{e18}
ds^2=-\f{\r^2\D_\th\D_r}{\S^2}dt^2+\f{\r^2}{\D_r}dr^2+\f{\r^2}{\D_\th}d\th^2
        +\l_{mn}(d\phi^m+N^mdt)(d\phi^n+N^ndt),
\ee
with $\phi^m=(\phi,\chi)$, $N^m=(N^\phi,N^\chi)$ and
\be
\l_{mn}=\lt(\ba{cc}
h_{\phi\phi}+A_1^2 & A_1 \\
A_1 & 1
\ea\rt).
\ee

From the above metric, besides the J-picture parameters~(\ref{e23}) we can also get the Q-picture parameters
\bea
&&f_1^Q|_{r=r_+}=\f{\sr{\r_+^2V}}{r_+^2+a^2}, ~~~
f_2^Q|_{r=r_+}=\sr{\f{\r_+^2}{V}}, ~~~
f_3^Q|_{r=r_+}=\f{q_e(r_+^2-a^2)}{(r_+^2+a^2)^2}, \nn\\
&&f^Q=m=\f{q_er_0^2(r_+^2-a^2)}{(r_+^2+a^2)^2}.
\eea
Therefore, considering the only change  from the function $f$, we find that
\bea \label{qp}
&&c_L^Q=\f{6m(r_+^2+a^2)}{\Xi}
       =\f{6q_e(r_+^2-a^2)(1-r_+^2/\ell^2)}
        {(1-a^2/\ell^2)(1+6r_+^2/\ell^2-3r_+^4/\ell^4-q^2/\ell^2)}, \nn\\
&&T_L^Q=\f{1}{2\pi m}
       =\f{(r_+^2+a^2)(1+6r_+^2/\ell^2-3r_+^4/\ell^4-q^2/\ell^2)}
      {2\pi q_e (r_+^2-a^2)(1-r_+^2/\ell^2)}.
\eea
This agrees exactly with the result found in~\cite{Chen:2010jj}. In the $a\ra0$ limit, we get the central charge and the temperature of extremal RN-AdS-dS black hole, in match with~\cite{Hartman:2008pb}. In the limit $1/\ell\ra0,q_m\ra0$, we have
\be \label{qp1}
c_L^Q=6Q^3,~~~T_L^Q=\f{r_+^2+a^2}{2\pi Q^3}.
\ee
Different from J-picture, the Q-picture is defined completely with respect to the rotation in $\chi$ direction, or equivalently with respect to the $U(1)$ gauge symmetry.

Note that from the uplifted 5D metric~(\ref{e18}), we can re-derive the result of the J-picture~(\ref{jp}) firstly derived from the metric~(\ref{j2}). Unlike~(\ref{j2}), the metric~(\ref{e18}) is the exact solution of the 5D Einstein-Hilbert action, and so the derivation here is an exact one. This somehow justifies the claim in~\cite{Hartman:2008pb} that the gauge field makes no contribution to the central charge.

\subsection{General picture}

Now we have the J- and Q-pictures, then according to Section 3, we still have the novel pictures generated by the $SL(2,\mb Z)$ modular group. Written explicitly, we have
\bea \label{gp}
&& c_L^{\phi'}=\f{6(ak+bm)(r_+^2+a^2)}{\Xi}, ~~~
T_L^{\phi'}=\f{1}{2\pi(ak+bm)},\nn\\
&&c_L^{\chi'}=\f{6(ck+dm)(r_+^2+a^2)}{\Xi}, ~~~
T_L^{\chi'}=\f{1}{2\pi(ck+dm)},
\eea
with the parameters
\be
\left(
  \begin{array}{cc}
    a & b \\
    c & d \\
  \end{array}
\right)
\in SL(2,\mb Z).
\ee
In the limit $1/\ell\ra0,q_m\ra0$, for the Kerr-Newman black hole we have
\bea
&& c_L^{\phi'}=6(2aJ+bQ^3), ~~~ T_L^{\phi'}=\f{r_+^2+a^2}{2\pi(2aJ+bQ^3)}, \nn\\
&& c_L^{\chi'}=6(2cJ+dQ^3), ~~~ T_L^{\chi'}=\f{r_+^2+a^2}{2\pi(2cJ+dQ^3)}.
\eea

\subsection{Near horizon geometry}

We may reproduce the above pictures by working on the near horizon geometry. Using the similar trick of obtaining the NHEK geometry, the near horizon geometry of an extreme Kerr-Newman-AdS-dS black hole can be written as~\cite{Hartman:2008pb}
\be \label{j4}
ds^2=\G(\th)\lt[ -r^2dt^2+\f{dr^2}{r^2}+\a(\th)d\th^2 \rt]+\g(\th)(d\phi+krdt)^2,
\ee
where
\bea
&&  \G(\th)=\f{\r_+^2r_0^2}{r_+^2+a^2},
   ~~~\a(\th)=\f{r_+^2+a^2}{\D_\th r_0^2},  \nn\\
&&  \g(\th)=\f{\D_\th(r_+^2+a^2)^2\sin^2\th}{\r_+^2\Xi^2}.
\eea
The near horizon gauge field is
\be
A=A_2(d\phi+krdt)+mrdt,
\ee
with
\be A_2=\f{1}{\r_+^2\Xi} \lt[ q_er_+a\sin^2\th+q_m(r_+^2+a^2)\cos\th \rt]. \ee
Note that we have used a different gauge from~\cite{Hartman:2008pb}, and it can be verified easily that there is just a constant discrepancy between the two gauges.

Similarly such a geometry could be taken as an extremal black hole with the horizon being at $r=0$. The horizon area is calculated to be
\be
\ma A=2\pi\int^\pi_0d\th\sr{\G(\th)\a(\th)\g(\th)}=\f{4\pi(r_+^2+a^2)}{\Xi}.
\ee
And from the metric we can read out
\be
f_1^J=f_2^J=\sr{\G(\th)},~~~f^J=f_3^J=k,
\ee
From~(\ref{e4}) and~(\ref{e5}), we  get the same central charge and temperature as~(\ref{jp}) in the J-picture.

To get the Q-picture, we uplift the metric~(\ref{j4}) to five dimensions
\be
ds_5^2=\G(\th)\lt[ -r^2dt^2+\f{dr^2}{r^2}+\a(\th)d\th^2 \rt]+\g(\th)(d\phi+krdt)^2+(d\chi+A)^2,
\ee
where $\chi\sim\chi+2\pi$. We recast it into the form
\be
ds_5^2=\G(\th)\lt[ -r^2dt^2+\f{dr^2}{r^2}+\a(\th)d\th^2 \rt]+\g_{mn}(d\phi^m+k^mrdt)(d\phi^n+k^nrdt),
\ee
where we have defined $k^m=(k,m)$ and
\be
\g_{mn}=\lt(\ba{cc}
\g(\th)+A_2^2 & A_2 \\
A_2 & 1
\ea\rt).
\ee
From the metric we can re-derive the J-picture, as well as the Q-picture parameters
\be
f_1^Q=f_2^Q=\sr{\G(\th)},~~~f^Q=f^Q_3=m.
\ee
Thus we can reproduce the results~(\ref{qp}) in the Q-picture. With the $SL(2,\mb Z)$ coordinate transformation, we can also obtain the same general picture as~(\ref{gp}).

\section{Higher-dimensional Kerr-AdS-dS black hole}

The metric of five-dimensional Kerr-AdS-dS black holes was obtained by Hawking, Hunter and Taylor-Robinson~\cite{Hawking:1998kw}, which generalizes the Ricci-flat rotating black hole of Myers and Perry~\cite{Myers:1986un}. As we see in the previous section, in the stretched horizon formalism, it does not matter whether we start from the  metrics of extremal black holes or the metrics of their near horizon geometries, as both lead to the same pictures. It turns out to be much easier to work on the near horizon geometry. For an extremal five-dimensional Kerr-AdS-dS black hole, its near horizon geometry could be described by the following metric~\cite{Lu:2008jk}
\bea
&&ds^2=\f{\r_+^2}{V}\lt( -r^2dt^2+\f{dr^2}{r^2} \rt)+\f{\r_+^2d\th^2}{\D_\th}
     +\f{\D_\th\sin^2\th}{\r_+^2}\lt(  \f{2a(r_+^2+b^2)}{r_+V}rdt
                                       +\f{r_+^2+a^2}{\Xi_a}d\phi_1   \rt)^2 \nn\\
&&\phantom{ds^2=}+\f{\D_\th\cos^2\th}{\r_+^2}\lt(   \f{2b(r_+^2+a^2)}{r_+V}rdt
                                       +\f{r_+^2+b^2}{\Xi_b}d\phi_2   \rt)^2 \\
&&\phantom{ds^2=}+\f{1+r_+^2/\ell^2}{r_+^2\r_+^2} \lt(   \f{2ab\r_+^2}{r_+V}rdt
                                       +\f{b(r_+^2+a^2)\sin^2\th}{\Xi_a}d\phi_1
                                       +\f{a(r_+^2+b^2)\cos^2\th}{\Xi_b}d\phi_2  \rt)^2,\nn
\eea
where
\bea
&& \r_+^2=r_+^2+a^2\cos^2\th+b^2\sin^2\th, ~~~V=4+4(3r_+^2+a^2+b^2)/\ell^2, \nn\\
&& \D_\th=1-\f{a^2}{\ell^2}\cos^2\th-\f{b^2}{\ell^2}\sin^2\th,
  ~~~\Xi_a=1-a^2/\ell^2,~~~\Xi_b=1-b^2/\ell^2.
\eea
The physical mass and angular momenta are
\be
M_{ADM}=\f{3\pi M}{4\Xi_a\Xi_b},~~~
J_{\phi_1}=\f{\pi Ma}{2\Xi_a^2},~~~
J_{\phi_2}=\f{\pi Mb}{2\Xi_b^2},
\ee
with
\be
M=\f{(r_+^2+a^2)^2(r_+^2+b^2)^2}{2r_+^4(2r_+^2+a^2+b^2)}.
\ee
The horizon area is
\be \ma A=\f{2\pi^2(r_+^2+a^2)(r_+^2+b^2)}{r_+\Xi_a\Xi_b}. \ee

There are two rotating Killing vectors in the geometries, just like in the uplifted Kerr-Newman black hole. It is thus expected that there are different dual CFT pictures to describe the black hole. We rewrite the above metric in the form
\be
ds^2=\f{\r_+^2}{V}\lt( -r^2dt^2+\f{dr^2}{r^2} \rt)+\f{\r_+^2d\th^2}{\D_\th}+q_{mn}e^me^n,
\ee
where we have defined
\bea
&& q_{11}=\f{\sin^2\th}{\r_+^2}\lt( 1-\f{a^2}{\ell^2}\cos^2\th+\f{b^2}{r_+^2}\sin^2\th \rt),\nn\\
&& q_{22}=\f{\cos^2\th}{\r_+^2}\lt( 1+\f{a^2}{r_+^2}\cos^2\th-\f{b^2}{\ell^2}\sin^2\th \rt),\nn\\
&& q_{12}=q_{21}=\f{ab(1+r_+^2/\ell^2)\sin^2\th\cos^2\th}{r_+^2\r_+^2},\nn\\
&& e^1=\f{2a(r_+^2+b^2)}{r_+V}rdt+\f{r_+^2+a^2}{\Xi_a}d\phi_1, \nn\\
&& e^2=\f{2b(r_+^2+a^2)}{r_+V}rdt+\f{r_+^2+b^2}{\Xi_b}d\phi_2.
\eea
We can then read the $J_1$ picture parameters
\be
f_1^{\phi_1}=f_2^{\phi_1}=\f{\r_+}{\sr V},~~~f^{\phi_1}=f_3^{\phi_1}=\f{2a\Xi_a(r_+^2+b^2)}{r_+V(r_+^2+a^2)}.
\ee
The $J_1$ picture turns out to have the central charge and the temperature
\be
c_L^{\phi_1}=\f{6\pi a(r_+^2+b^2)^2}{r_+^2V\Xi_b},~~~
T_L^{\phi_1}=\f{r_+V(r_+^2+a^2)}{4\pi a\Xi_a(r_+^2+b^2)}.
\ee
In the similar way we can get the $J_2$ picture
\be
c_L^{\phi_2}=\f{6\pi b(r_+^2+a^2)^2}{r_+^2V\Xi_a},~~~
T_L^{\phi_2}=\f{r_+V(r_+^2+b^2)}{4\pi b\Xi_b(r_+^2+a^2)}.
\ee
The above results are exactly the same as the ones obtained in~\cite{Lu:2008jk}.

Similar to the four-dimensional Kerr-Newman case, we can get more general picture. As the analysis is quite similar, we do not
present the details here. In a general picture,  the central
charge and the temperature are respectively
\bea
&& c_L^{\phi'_1}=a c_L^{\phi_1} + b c_L^{\phi_2}, ~~~ T_L^{\phi'_1}=\f{1}{a/T_L^{\phi_1}+b/T_L^{\phi_2}} ,\nn\\
&& c_L^{\phi'_2}=c c_L^{\phi_1} + d c_L^{\phi_2}, ~~~ T_L^{\phi'_1}=\f{1}{c/T_L^{\phi_1}+d/T_L^{\phi_2}},
\eea
with
\be
\lt(\ba{ll} a&b\\c&d \ea\rt)\in SL(2,\mb Z).
\ee


The discussion above can be generalized to higher dimensions.
The metric of near horizon geometry of the  higher dimensional extreme Kerr-AdS-dS black holes were given in~\cite{Lu:2008jk}. For even dimensions $D=2n$, there are $n-1$ rotation angles, and for each angle we have a CFT dual
\bea
&&c_i=\f{3a_ir_+\ma A_{D-2}}{\pi V} \prod_{j\neq i}\f{(r_+^2+a_j^2)^2}{\Xi_j}, \nn\\
&&T_i= \f{V(r_+^2+a_i^2)}{4 \pi a_i \Xi_i r_+ \prod_{j \neq i} (r_+^2+a_j^2) },
\eea
and
\be S_{BH}=\f{\pi^2}{3}c_iT_i~~~\textrm{for each}~i, \ee
where $V$ is a constant introduced in~\cite{Lu:2008jk}, and $\ma A_{D-2}=2\pi^{(D-1)/2}/\G[(D-1)/2]$ is the volume of the unit (D-2)-sphere.

The CFT dual could be derived with respect to particular angular directions. Also we can make a $SL(n-1,\mb Z)$ rotation, define new angular coordinates and analyze the CFT duals with respect to the new angles. We define the $SL(n-1,\mb Z)$ matrix $\L^{m'}_{\phantom{m'}m}$, the CFT dual to the angle $\phi^{m'}=\L^{m'}_{\phantom{m'}m}\phi^m$ turns out to be
\be \label{e16}
c_L^{m'}=\L^{m'}_{\phantom{m'}m} c_L^m, ~~~ T_L^{m'}=\f{1}{\L^{m'}_{\phantom{m'}m}/T_L^m}.
\ee

For odd dimensions $D=2n+1$, there are $n$ rotation angles, and we have the discrete CFT duals with
\bea
&&c_i=\f{3a_i\ma A_{D-2}}{\pi V r_+^2} \prod_{j\neq i}\f{(r_+^2+a_j^2)^2}{\Xi_j}, \nn\\
&&T_i= \f{V r_+(r_+^2+a_i^2)}{4 \pi a_i \Xi_i \prod_{j \neq i} (r_+^2+a_j^2) }.
\eea
Similar to the even dimensions, we can use a $SL(n,\mb Z)$ matrix $\L^{m'}_{\phantom{m'}m}$ to have the general CFT duals.

\section{Equivalence to ASG formalism}

 In this section we show  the equivalence of the stretched horizon formalism to the asymptotic symmetry group (ASG) formalism, or sometimes called BBC (Barnich-Brandt-Compere) formalism~\cite{Barnich:2001jy,Barnich:2007bf}.
 In the ASG formalism we have to firstly propose a set of consistent boundary conditions for the near horizon geometry, and get the corresponding asymptotic Killing vector. The ASG is defined as the set of allowed symmetry transformations modulo the set of trivial ones. The ASG forms a set of Witt algebra, which upon quantization  becomes the Virasoro algebra with the central charge. The details can be found in~\cite{Guica:2008mu}.

For the 4D case, the analysis starts from the near horizon geometry~(\ref{z1}). In the coordinates $(t,r,\th,\phi)$ and with the boundary conditions as proposed in~\cite{Guica:2008mu}
\be
h_{\m\n}=\ma O \lt( \ba{ccccc}
r^2&1/r^2&1/r&1\\
   &1/r^3&1/r^2&1/r\\
     &&1/r&1/r\\
      &&&1\\
\ea \rt),
\ee
the ASG turns out to be
\be
\z=\e(\phi)\p_\phi-\e'(\phi)r\p_r,
\ee
with $\e(\phi)$ being arbitrary function of $\phi$. Since the coordinate $\phi$ has the period $\phi \sim \phi+2\pi$, we can expand $\e_n(\phi)=-e^{-in\phi}$ and get
\be
\z_n=-e^{-in\phi}(\p_\phi+inr\p_r),
\ee
where $n$ is an arbitrary integer. These vector fields generate a Witt algebra
\be
i[\z_m,\z_n]=(m-n)\z_{m+n}.
\ee
The conserved charges $Q_\z$ generated by these diffeomorphisms are defined as the integral over a spatial slice $\p\S$
\be
Q_\z=\f{1}{8\pi G}\int_{\p\S}k_\z[h,g],
\ee
where the 2-form $k_\z[h,g]$ is defined for a perturbation $h_{\m\n}$ around the background metric $g_{\m\n}$ by
\be
k_\z[h,g]=\f{1}{2} \lt[ \z_\n\na_\m h-\z_\n\na_\s h_\m^{\phantom\m\s}+
           \z_\s\na_\n h_\m^{\phantom\m\s}+\f{1}{2}h\na_\n\z_\m
           +\f{1}{2} h_{\n\s} (\na_\m \z^\s -\na^\s\z_\m)  \rt] *(dx^\m \wedge dx^\n) ,
\ee
where $*$ denotes the Hodge dual. The Dirac bracket algebra of the charges contains a central term
\be
\{ Q_{\z_m},Q_{\z_n} \}_{DB}=Q_{[\z_m,\z_n]}+\f{1}{8\pi G}\int_{\p\S}k_{\z_m}
                                               [\ma L_{\z_n}g,g].
\ee
The corresponding commutator algebra is given by
\be
[L_m,L_n]=(m-n)L_{m+n}+\f{c_L}{12}(m^3+\ma B m)\d_{m+n},
\ee
with the central charge given by
\be \label{e17}
\f{1}{8\pi G}\int_{\p\S}k_{\z_m}[\ma L_{\z_n}g,g]=\f{-i c_L}{12}(m^3+\ma B m)\d_{m+n},
\ee
and $\ma B$ is some possible unimportant constant. Explicit calculation shows that the central charge is
\be c_L=\f{3f}{G}\int^\pi_0 d\th \sr{\G(\th)\a(\th)\g(\th)}=\f{3f \ma A}{2\pi G},\ee
being the same with~(\ref{e4}). Therefore we prove that the stretched horizon formalism is completely equivalent to the formalism in~\cite{Guica:2008mu,{Hartman:2008pb}}, based on the asymptotical symmetry analysis of near horizon geometry.

Similarly, in the 5D case, for the near horizon geometry~(\ref{e22}) in the coordinates $(t,r,\th,\phi,\chi)$, we propose the boundary conditions~\cite{Compere:2009dp}
\be \label{bj}
h_{\m\n}=\ma O \lt( \ba{ccccc}
r^2&1/r^2&1/r&1&r\\
   &1/r^3&1/r^2&1/r&1/r^2\\
     &&1/r&1/r&1/r\\
      &&&1&1\\
       &&&&1/r
\ea \rt),
\ee
and then we have the asymptotic symmetry
\be
\z^{(\phi)}=\e(\phi)\p_\phi-\e'(\phi)r\p_r.
\ee
Expanding $\e_n(\phi)=-e^{-in\phi}$, we get the generators
\be \label{vj}
\z^{(\phi)}_n=-e^{-in\phi}(\p_\phi+inr\p_r),
\ee
which satisfy the commutation relation of a Witt algebra in $\phi$-picture.

Moreover, we may propose another set of consistent boundary conditions
\be \label{bq}
h_{\m\n}=\ma O \lt( \ba{ccccc}
r^2&1/r^2&1/r&r&1\\
   &1/r^3&1/r^2&1/r^2&1/r\\
     &&1/r&1/r&1/r\\
      &&&1/r&1\\
       &&&&1
\ea \rt),
\ee
which have the asymptotic symmetry
\be
\z^{(\chi)}=\e(\chi)\p_\chi-\e'(\chi)r\p_r.
\ee
Expanding $\e_n(\chi)=-e^{-in\chi}$, we get the generators of the Witt algebra in $\chi$-picture
\be \label{vq}
\z^{(\chi)}_n=-e^{-in\chi}(\p_\chi+inr\p_r).
\ee

Note that the two Virasoro are independent and mutually commutable. Using the formula~(\ref{e17}), we obtain the  central charges in the $\phi$- and $\chi$-picture respectively
\bea
&&c_L^\phi=\f{6\pi f^\phi}{G}\int^\pi_0 d\th \sr{\G(\th)\a(\th)\lt( \g_{\phi\phi}(\th)\g_{\chi\chi}(\th)-\g_{\phi\chi}(\th)^2 \rt)}
          =\f{3 f^\phi\ma A}{2\pi G},\nn\\
&&c_L^\chi=\f{6\pi f^\chi}{G}\int^\pi_0 d\th \sr{\G(\th)\a(\th)\lt( \g_{\phi\phi}(\th)\g_{\chi\chi}(\th)-\g_{\phi\chi}(\th)^2 \rt)}
          =\f{3 f^\chi\ma A}{2\pi G},\nn
\eea
which are the same as~(\ref{e19}) and~(\ref{e20}).

It is remarkable that the stretched horizon formalism always gives rise to the same CFT duals as the ones suggested by ASG of near horizon geometry. Note that in the stretched horizon formalism, the constraint or the boundary conditions on the metric components are much less. Actually there is only one constraint~(\ref{shift}) on the shift vector $N^i$. On the contrary, in the ASG formalism, one has to know the falloff conditions on all the metric components of near horizon geometry. In this sense, the stretched horizon formalism is easier to work with. Nevertheless, the common point in both formalism is that the $U(1)$ isometry group get enhanced to a Virasoro algebra with non-vanishing central charge. It is this feature guiding us to find more general dual pictures for the extreme black holes with more than one $U(1)$ symmetry.
It is possible to study the novel pictures generated by T-duality group $SL(n,\mb Z)$ in the ASG formalism. However, it is in the stretched horizon
formalism, the novel pictures could be read out easily.

\section{Conclusions and discussions}

In this paper, we investigated the holographic descriptions of extreme black holes in the stretched horizon formalism.
We developed the original derivation of Kerr/CFT correspondence in~\cite{Carlip:2011ax} to more general extremal backgrounds.
It turns out that for the dual CFT the central charges is always proportional to the entropies of the black holes and the temperatures
is given by the value of a function at the horizon.
We not only reproduced successfully the well-established holographic pictures of BTZ, RN, Kerr-Newman-AdS-dS and Kerr black holes in higher
dimensional spacetime, but also proposed novel pictures for the extremal black holes with no less than two $U(1)$ symmetries. More precisely we showed
that for the black hole with $n$ $U(1)$ symmetries, there are  not only CFT duals with the central charge $c_L^m$ and the temperature $T^m_L$ for each $U(1)$ symmetry, but also more CFT duals generated by
$SL(n,\mb Z)$ group with the central charges and the temperatures
\bea
c_L^{m'}=\L^{m'}_{\phantom{m'}m}c_L^m,\hs{3ex}\f{1}{T_L^{m'}}=\L^{m'}_{\phantom{m'}m}\f{1}{T_L^{m}},
\eea
with $\L^{m'}_{\phantom{m'}m}\in SL(n,\mb Z)$.

In our study, we noticed that it does not matter whether we start from the usual extremal black hole geometries or from their near horizon geometries, both of which lead to the same dual pictures in the stretched horizon formalism. We proved  this equivalence in four and five dimensions and expected it in other dimensions. Besides, for the four- and five-dimensional stationary extremal black holes, we proved the equivalence between the stretched horizon formalism and usual ASG analysis of near horizon geometry. Such equivalence should be true in other dimensions. However, the two formalisms are different. In the stretched horizon formalism, one obtains the boundary asymptotic symmetry just from the constraint on the shift vector $N^i$~(\ref{shift}). On the other hand, in the analysis of ASG of near horizon geometry  the falloff conditions on all of the metric perturbation have to be taken into account. Nevertheless, in both approach, it is the $U(1)$ symmetry of the rotation which get enhanced to a Virasoro algebra with non-vanishing central charge. It would be interesting to understand the relations between two formalisms better.

In~\cite{Carlip:2011ax} Carlip also tried to extend the method to non-extremal Kerr black holes, but found that the temperature and the central charge behave poorly at the horizon and only half of the Bekenstein-Hawking entropy can be reproduced. Generally we believe that for a non-extremal black hole there is a non-chiral CFT dual, with both non-vanishing left and right central charges and temperatures, from which the Cardy formula gives the full Bekenstein-Hawking entropy
\be S=\f{\pi^2}{3}(c_LT_L+c_RT_R). \ee
The naive generalization of stretched horizon formalism to non-extremal cases yields many problems. It is still an open issue how to derive the central charges of both sectors in a convincing way.

One interesting effort in setting up the non-extremal Kerr/CFT correspondence is to use the idea of hidden conformal symmetry~\cite{Castro:2010fd}. Assuming the form of the central charges is kept invariant, one may read out the dual temperatures from the conformal coordinates. Such a strategy works well for various black holes. Especially for a four-dimensional Kerr-Newman black hole, both the J-picture and Q-picture have been set up in this framework~\cite{HiddenSymmetry,Chen:2010bh,Chen:2010jj,Chen:2010ywa}. It would be nice to check if the general dual picture proposed in this paper make sense or not for non-extremal black holes, with the help of hidden conformal symmetry. In fact we have found that the general picture can be set up from the hidden conformal symmetry of the generic non-extreme black holes~\cite{Chen:2011kt}.

Another interesting issue concerns the possible relation of a general dual picture to string theory. In~\cite{Guica:2010ej}, it was proposed that the J-picture and Q-picture of the Kerr-Newman black hole could be related to the long and short string pictures. It would be interesting to find a string picture for the general case. In our study, we showed that for higher dimensional Kerr and Kerr-Newman black holes with $n$ $U(1)$ symmetries, there exist a  transformation $SL(n,\mb Z)$ generating the novel CFT duals. It would be nice to find a construction in string theory for these black holes.

\vspace*{12mm}

\noindent
 {\large{\bf Acknowledgments}}

The work was in part supported by NSFC Grant No. 10975005. We would like to thank Chiang-mei Chen for stimulating discussion on possible relation between J- and Q-pictures for Kerr-Newman black holes.

\vspace*{5mm}


\end{document}